\documentclass{article}
\usepackage{amsmath}
\usepackage{graphicx}
\usepackage[dvips]{epsfig}

\begin{document}

%********************************************************************
%  TITLE PAGE
%********************************************************************

\begin{flushright}
   MZ-TH/02-28  \\
   November 2002 \\
\end{flushright}

\vspace*{1cm}

\begin{center}
\textbf{\large Charged pion pair production by two photons from a chiral sum
rule } \vspace{10mm} \\[0pt]
N.\ F.\ Nasrallah%\footnote{ } 
\\[8mm]
\textit{Faculty of Science, Lebanese University, Tripoli, Lebanon \\[12mm]
}and K.~Schilcher%\footnote{ } 
\\[8mm]
\emph{Institut f\"{u}r Physik, Johannes-Gutenberg-Universit\"{a}t,\\[0pt]
Staudinger Weg 7, D-55099 Mainz, Germany}
\end{center}

\bigskip

\textbf{Abstract: } The cross-section for a charged pion pair production by
two photons is evaluated by using the low energy expression previously
obtained from current algebra and PCAC which involves an integral over
vector and axial-vector spectral functions. Data on the latter obtained from 
$\tau $-decay as measured by the ALEPH collaboration is then inserted in the
integral which is appropriately modified in order to eliminate contributions
near the cut in the duality contour integral. The ex\-pe\-ri\-men\-tal
behavior of the cross-section for $\gamma \gamma \rightarrow \pi ^{+}\pi ^{-}
$ is well reproduced at low energies.

\vspace{12mm}

\newpage

%%%%%%%%%%%%%%%%%%%%%%%%%%%%%%%%%%%%%%%%%%%%%%%%%%%%%%%%%%%%%%%%%%%%%%%
%\section{}
\vspace{2cm}

Two photon processes for charges pion pair production have recently become a
subject of experimental investigation by the use of $e^{+}e^{-}$ colliding
beams. Measurements of the cross-section for the two-photon production of
charged pion and kaon pairs have been performed by the Mark II\cite{MarkII}
and TPC/Two-Gamma\cite{TPC} collaborations and, more recently, by the CLEO%
\cite{cleo} collaboration

On the theoretical side pion pair production by photons at low energies may
is best described by of chiral perturbation theory (CHPT)\cite{chpt}. The
Born term dominates the behavior of the cross-section at low energies.
Next-to-leading corrections (one loop graphs) change the result very little%
\cite{Bijcor}. Two-loop contributions to the amplitude have recently been
presented by B\"{u}rgi\cite{Bi}. As is well known, this elaborate
calculation involves new unknown coupling constants (3 in this case), which
were estimated in\cite{Bi} by use of resonance saturation.\newline
An alternative approach based on dispersion theory has also reached a
considerable degree of sophistication \cite{pennington}

In this note we want to present a simple analysis of the process $\gamma
\gamma \rightarrow \pi ^{+}\pi ^{-}$ based on a chiral sum rule proposed by
Terazawa \cite{Terazawa1}. This sum rule was an early application of
current-algebra and the partially conserved axial-vector current (PCAC)
hypothesis. The method of ref.\cite{Terazawa1} gives an elegant expression
of the cross-section for $\gamma \gamma \rightarrow \pi ^{+}\pi ^{-}$ in
terms of vector and axial-vector spectral function integrals. An update of
the calculation presented in ref.\cite{Terazawa1} appeared recently\cite%
{Terazawa} in which $\rho $ and $A_{1}$ meson pole dominance of the spectral
function integrals was used and the result was expressed in terms of an
effective vector meson mass $m_{v}$ which was determined by a best fit to be 
$m_{v}\simeq 1.4GeV$ , a value somewhat larger than the $\rho $-mass of $%
0.77GeV$. This result is just a manifestation of the fact that the
assumption of vector meson dominance is notoriously unreliable if, as is the
case in Terazawa's sum rule, the small difference of two large spectral
functions enters.

In the meantime the ALEPH\cite{Aleph} collaboration has obtained detailed
and precise experimental information about the vector and axial-vector
spectral function from $\tau $ decay, which can be used in Terazawa's sum
rule. We shall argue that the sum rule has to be appropriately modified in
order to minimize the contribution of the contour near the cut in the
duality integral.

The PCAC hypothesis and current algebra yield the following expression for
the differential cross-section of the reaction $\gamma \gamma \rightarrow
\pi ^{+}\pi ^{-}$ in terms of the scattering angle $\theta $

\begin{equation}
\frac{d\sigma }{d(\cos \theta )}=\frac{\pi \alpha ^{2}}{s}(1-\frac{4m_{n}^{2}%
}{s})^{1/2}F^{2}(s/2)  \label{eq1}
\end{equation}%
where $s$ is the invariant mass-squared of the photoproduced pion pair and $%
F(s)$ is the pion structure function

\begin{equation}
F(s)=\frac{1}{4\pi f_{\pi }^{2}}\int^{\infty }dt\frac{t}{(s+t)}(\rho
_{V}(t))-\rho _{A}(t))  \label{eq2}
\end{equation}%
where $f_{\pi }=92.4$ MeV is the pion decay constant and where $\rho _{V}$
and $\rho _{A}$ are the spectral functions of the vector and axial-vector
currents respectively. We subdivide the interval into two regions, one from
threshold to $R=3GeV^{2}$, where the ALEPH data are measured with good
precision, and a tail from $R$ to $\infty $. As there is no singularity for $%
\left\vert s\right\vert >R$, eq.(\ref{eq2}) can be written by Cauchy's
theorem as%
\begin{eqnarray}
F(s) &=&\frac{1}{4\pi f_{\pi }^{2}}\int^{R}dt\frac{t}{(s+t)}(\rho
_{V}(t))-\rho _{A}(t))  \label{eq2a} \\
&&-\frac{1}{2i}\frac{1}{4\pi f_{\pi }^{2}}\oint dt\frac{t}{(s+t)}(\Pi
_{V}(t))-\Pi _{A}(t))\ ,  \notag
\end{eqnarray}%
where\ $\rho (t)={\rm Im}\Pi (t)$ and the second integral\ is extended over
a circle of radius $R$. For the first integral the ALEPH data can be used,
and in the second integral one is tempted to use the QCD expressions for $%
\Pi _{V,A}$. From a study of the Weinberg sum rules \cite{DS} it is known
that the QCD expressions cannot be trusted in the vicinity of the positive
real axis for $t\simeq R$.

In order to minimize the uncertainty introduced by the inadequate QCD
expression near the positive real axis we make use of the second Weinberg
sum rule\cite{Weinberg} which holds in QCD

\begin{equation}
\frac{1}{4\pi f_{\pi }^{2}}\int^{\infty }dt\quad t(\rho _{V}(t))-\rho
_{A}(t))=0\ .  \label{eq3}
\end{equation}%
This allows us to write

\begin{eqnarray}
F(s) &=&\frac{1}{4\pi ^{2}f_{\pi }^{2}}\int^{R}dt\quad t(\frac{1}{s+t}-\frac{%
1}{s+R})[\rho _{V}(t)-\rho _{A}(t)]  \label{eq4} \\
&&+\frac{1}{4\pi ^{2}f_{\pi }^{2}}\int_{R}^{\infty }dt\quad t(\frac{1}{s+t}-%
\frac{1}{s+R})[\rho _{V}^{QCD}(t)-\rho _{A}^{QCD}(t)]
\end{eqnarray}

The second integral above extends to infinity but the integrand vanishes
rapidly with increasing $t$. The contribution of the tail can be estimated
by using the QCD result \cite{chetyrkin}%
\begin{equation}
\rho _{V}^{QCD}-\rho _{A}^{QCD}=\frac{32\pi }{9}\alpha _{s}(t)\left\langle
\alpha _{s}(\overline{q}q)^{2}\right\rangle \frac{1}{t^{3}}  \label{2.1}
\end{equation}%
Using the standard value $\left\langle \alpha _{s}(\overline{q}%
q)^{2}\right\rangle \sim 1.0\times 10^{-4}GeV^{6}$ shows that this
contribution is negligible.

In order to compare with the experiment we have to make use of the fact that
the measured values of the scattering angle are limited to $|\cos \theta
|\leq .6$. The total cross-section for the process $\gamma \gamma
\rightarrow \pi ^{+}\pi ^{-}$ is thus

\begin{equation}
\sigma (|\cos \theta |\leq .6)=\frac{\pi \alpha ^{2}}{s}(1-\frac{4m_{\pi
}^{2}}{s})^{1/2}|F(s/2)|^{2}\times 1.2  \label{eq5}
\end{equation}%
\begin{figure}[h] 
\unitlength 1mm
\begin{picture}(140,70)
\put(20,-10){\epsfig{file=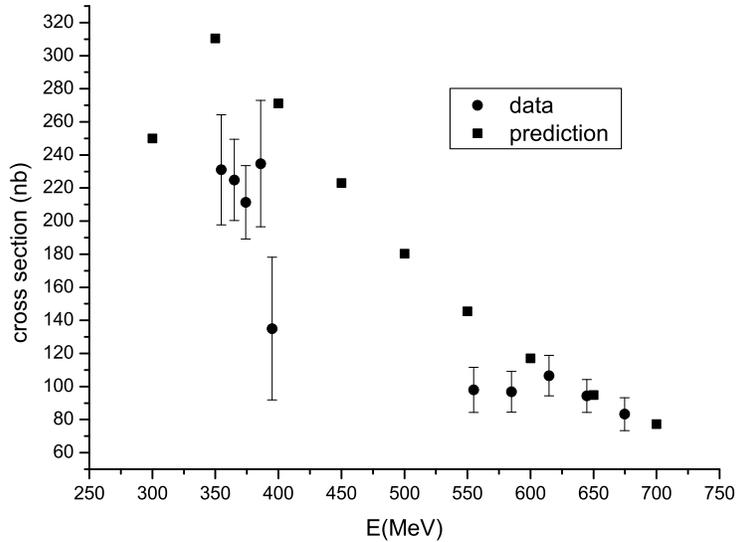,width=11cm}}
\end{picture}
\caption{Predictions for the $\protect\gamma \protect\gamma \rightarrow
  \protect\pi ^{+}\protect\pi ^{-}$ cross section versus the data
}
\label{fig1}
\end{figure}

This expression for $\gamma \gamma \rightarrow \pi ^{+}\pi ^{-}$ cross
section the is plotted in Fig. (1) where the data are also represented.
From the figure we see that a simple analysis based on the chiral sum
rule of Terazawa combined with the ALEPH data on vector and axial-vector
spectral functions yields a consistent description of the low energy
behavior of the photo production of charged pions.

%%%%%%%%%%%%%%%%%%%%%%%%%%%%%%%%%%%%%%%%%%%%%%%%%%%%%%%%%%%%%%%%%%%%%%

%----------------------------------------------------------------------


\vspace{7mm}\noindent

\begin{thebibliography}{99}
\bibitem{MarkII} Mark II Collaboration, J. Boyer \emph{et al\/} \textit{%
Phys. Rev. Lett.} \textbf{56} (1986) 207

\bibitem{TPC} TPC/Two-Gamma Collaboration, H. Aihora \emph{et al\/} \textit{%
Phys. Rev. Lett.} \textbf{57} (1986) 404

\bibitem{cleo} CLEO Collaboration , J. Dominick \emph{et al\/} \textit{Phys.
Rev.} \textbf{D50} (1994) 3024

\bibitem{chpt} S. Weinberg, 1979 \textit{Physica} \textbf{A96} (1979) 327
;J. Gasser, H. Leutwyler, \textit{Am. Phys.} \textbf{\ 158} (1984) 142; J.
Gasser, H. Leutwyler, 1980 \textit{Nucl. Phys.} \textbf{\ B250,} 465;J.
Gasser, H. Leutwyler, \textit{Am. Phys.} \textbf{\ 235} (1994) 165

\bibitem{Bijcor} J. Bijnens, F. Cornet, \textit{Nucl. Phys.} \textbf{\ B296}
(1988) 557

\bibitem{Bi} U. B\"{u}rgi, Nucl.Phys.\textbf{B479} (1996) 392;

\bibitem{pennington} D. Morgan, M. R. Pennigton, Z. Phys. \textbf{C48}
(1990) 623; G. Mennessier, T. N. Truong, \textit{Phys. Lett. }\textbf{B177}
(1986) 195

\bibitem{Terazawa1} H. Terazawa, \textit{Phys. Rev. Lett.} \textbf{26 (}%
1971) 1207;

\bibitem{Terazawa} H. Terazawa, \textit{Phys. Rev.} \textbf{D51} (1995) R954

\bibitem{Aleph} The ALEPH Collaboration, R. Barate et al. \textit{Eur. Phys.
J.} \textbf{\ C76 (}1997), 15;\newline
1998 \textit{Nucl. Phys.} \textbf{C4} , 409

\bibitem{Weinberg} S. Weinberg, \textit{Phys. Rev. Lett.} \textbf{18 }%
(1967), 507

\bibitem{chetyrkin} L. V. Larin, V. P. Spiridonov, K. Chetyrkin Yad. Fiz. 
\textbf{44 }(1986) 1372 [Sov: J. Nucl. Phys.\textbf{44} (1986)892]

\bibitem{DS} C. A. Dominguez, K. Schilcher, Phys.Lett.\textbf{B448} (1999) 93
\end{thebibliography}
\end{document}